\documentclass[pre,twocolumn,superscriptaddress,preprintnumbers,aps,showpacs,amsmath,amssymb]{revtex4}

\usepackage[pdftex]{graphicx}
\usepackage{dcolumn}
\usepackage{bm}
\usepackage{color}
\usepackage{longtable}
\usepackage{amsmath}	

\usepackage{xr}
\newcommand{\pdiff}[2]{\frac{\partial #1}{\partial #2}}
\newcommand{\odiff}[2]{\frac{d #1}{d #2}}

\begin{document}

\preprint{}

\title{
Self-propulsion of an active polar drop
}


\author{Natsuhiko Yoshinaga }
\email[E-mail: ]{yoshinaga@tohoku.ac.jp}
\affiliation{
WPI - Advanced Institute for Materials Research, Tohoku University,
Sendai 980-8577, Japan
}
\affiliation{
MathAM-OIL, AIST, Sendai 980-8577, Japan
}



 \begin{abstract}
  We investigate the self-propulsive motion of a drop containing an active
  polar field.
  The drop demonstrates spontaneous symmetry breaking from a uniform
  orientational order into a splay or bend instability depending on the
  types of active stress, namely, contractile or extensile,
  respectively.
  We develop the analytical theory of the mechanism of this instability,
  which has been observed only in numerical simulations.
  We show that both contractile and extensile active stress result in
  the instability and self-propulsive motion.
  We also discuss asymmetry between contractile and extensile stress,
  and show that extensile active stress generates chaotic motion even
  under a simple model of the polarity field coupled with motion and
  deformation of the drop. 
 \end{abstract}


\maketitle

\section{Introduction}

Cell motility is a fascinating phenomenon in biological systems.
Depending on the cell type, cells may move in a fluid, on a substrate, in a tissue consisting of
other cells, or many other environments \cite{bray:2000}.
It is evident that this phenomenon is a result of the complex cooperation of
biomacromolecules and therefore it is difficult to understand its
mechanism.
Nevertheless, simple models may give an insight into the essential aspects of
the mechanism responsible for cell motility, and assist in the
classification of its
mechanism into several universality classes\cite{Camley:2017}.
This direction of research has been met with great success, for example,
for micro-organisms in a fluid, which is described by a squirmer
model\cite{Lighthill:1952,blake:1971c}.
The model imitates the motion of flagella and cilia on a cell body as an
effective slip boundary on a solid (but possibly deformable) body, and
all the complexity of propulsion falls into a functional form of the
effective slip boundary condition of a fluid dynamics problem.
Despite the reasonable generality of the model, the
problem is analytically tractable for a simple geometry, and for this
reason, extensive studies have been carried out to clarify that it can
reproduce realistic  motion near a wall, interaction between swimmers, and
collective behaviors\cite{Lauga:2009}.

The extension of the idea toward much more complex motility such as keratocyte
or amoeboid movement on a substrate is in its infancy.
Such motility is driven by the activity of a cytoskeleton, the mechanics
of which is
controlled by the collective behaviors of actin filaments and myosin
molecular motors.
The activity of a cytoskeleton falls into two classes: (i) actin
polymerization and (ii) contractility of actin filaments mediated by
molecular motors.
The latter may be classified into two sub-categories: (ii-a)
contractility on cortex on the surface of a cell and (ii-b)
contractility in cytosol in the bulk.
Recently, several models have been proposed to describe each mechanism
such as polarity-driven cell motility \cite{shao:2010,Ziebert:2012,Aranson:2015,Loeber:2015,Camley:2017b} for (i)
\footnote{
See also \cite{callan-jones:2008,BlanchMercader:2013} discussing another
mechanism of polarity-driven self-propulsion, namely,
deformation-induced motion.
}
, and motility induced by active stress \cite{Tjhung:2012,Recho:2013,Tjhung:2015} for (ii-b).
The surface contractility results in blebbing, which is considered as
another type of motility \cite{Charras:2008,Kapustina:2016}.

The main ingredients of these models are the field that describes
position and shape of the moving cell, and the polarity field that
represents orientation and its magnitude of actin filaments inside the cytoskeleton.
In this respect, they are regarded as active hydrodynamics confined in a
drop whose interface is moving by the internal force generated by active fluid\cite{Marchetti:2013}.
This combination of the polarity field and free-boundary problem is one
of the main obstacles to analytical treatment.
This is, in fact, another motivation for this study, namely, we would
like to compare the models of cell motility and self-propulsion of a chemically driven drop\cite{Yoshinaga:2017}.
In the latter model, a drop produces or consumes a concentration field
such as surfactants by chemical reaction, resulting in inhomogeneous
surface tension and motion of the drop through the Marangoni effect\cite{Yabunaka:2012,Yoshinaga:2012a,Yoshinaga:2014,Yoshinaga:2018b}.
This model shares the free-boundary problem of the drop with
the model of contractility-driven cell motility, but the polarity field
is replaced by a scalar field describing the concentration of chemical
molecules.
Owing to the simplicity of the scalar field, various analytical
treatments have been proposed\cite{Yabunaka:2016,Goff:2017,Seyboldt2018}.
The polarity field is, on the other hand, very difficult to treat
because of its vectorial nature.
 This drawback of its complexity may be compensated for by a richer structure in
 the model; as we will show, the active polar drop may demonstrate more
 various motion than the motion of chemically driven drops where only
 straight and helical motion have been observed.
 By analyzing the model with the polar field, we hope to clarify
 the similarity and differences between these models.

 Active polar or nematic fluids exhibit distinct features compared from
 simple isotropic fluid and also from conventional liquid crystal, even
 under unconfined systems.
First, the fluid may show topological defects (disclination), which plays
 an important role in its dynamics.
 Although the defects are also seen in conventional liquid crystals,
the motion of the defect is strongly
 influenced by its topological charge (sign); for example, only the
 $1/2$ disclination of the active nematic fluid is able to move spontaneously\cite{Pismen:2013,Giomi:2013,Giomi:2014}.
 Another important feature of the
 active fluid is that uniform orientation becomes unstable as activity
 increases.
 Eventually, it demonstrates turbulence-like flow containing a large
 number of topological defects\cite{Thampi:2014a}.
 This instability occurs by coupling between fluid flow and the polarity field.
It was first proposed theoretically in \cite{Simha:2002}, followed by detail
 theoretical studies\cite{Voituriez:2005,Edwards:2009} as well as numerical analysis \cite{Giomi:2012}.
Depending on the sign of the active stress, contractile and extensile
 stress results in splay and bend instability, respectively, in a flow-tumbling regime.
 On the other hand, in a flow-aligning regime for a rod-like shape, only the extensile stress
 is linearly unstable\cite{Edwards:2009}.

 Intuitively, the active polar drop uses the mechanism of this
 instability to move spontaneously\cite{Tjhung:2012}.
 Our claim is that this is true for the extensile stress, whereas for the contractile stress,
 the effect of the interface confining the active fluid inside the drop must be considered.
 We discuss how the interface gives rise to additional flow.
 We also show that after the instability, the distortion of the polarity
 field produces source dipole flow resulting in self-propulsion.

 In this work, we focus on active polar drops where the
motility is driven by bulk contractility
generated by active stress.
To the best of the author's knowledge, most previous studies have focused only on numerical
simulations in two \cite{Tjhung:2012,Giomi:2014a,Marth:2014,Fialho:2017} or three
dimensions \cite{Tjhung:2015}.
There have been scarce theoretical studies on active drops.
In \cite{Whitfield:2014}, the speed of an active polar drop is computed by
assuming an ansatz of the polar field \cite{Whitfield:2014}.
The flow field inside a thin film of a drop was analytically calculated in \cite{Khoromskaia:2015}.
 These works focused on fluid flow and the speed of a drop under a given
 polarity field designed for their analysis.
However, it is not clear how transition (bifurcation) occurs between
 the stationary state and the self-propelled state.
 The main focus in this work is to investigate the mechanism of the
 transition (drift bifurcation) using as a simple setting as possible.

\section{Model}

We consider an active polar field composed in a drop with a radius $R$
in two dimensions.
The model for polarity distribution, ${\bf P}({\bf x})$, is written as\cite{Tjhung:2012}
\begin{align}
\pdiff{{\bf P}}{t} 
+ \left(
\left(
{\bf v} + v_p {\bf P}
\right) \cdot \nabla
\right) {\bf P}
+ \boldsymbol{\omega} \cdot {\bf P}
&= \xi \boldsymbol{\kappa} \cdot {\bf P}
 - \frac{1}{\Gamma} \frac{\delta F}{\delta {\bf P}}
 \label{active.polar2.drop.model}
\end{align}
where the symmetric and anti-symmetric parts of the velocity gradient
tensors are denoted by $\kappa_{ij}({\bf x})$ and
$\omega_{ij}({\bf x})$, respectively.
They are explicitly given by the gradient of the velocity field ${\bf
v}({\bf x})$ as
\begin{align}
\kappa_{ij} 
&=
\frac{1}{2} \left(
\nabla_i v_j
+ \nabla_j v_i
\right)
\\
\omega_{ij}
&=
\frac{1}{2} \left(
\nabla_i v_j
- \nabla_j v_i
\right)
.
\end{align}
The term $\xi \boldsymbol{\kappa} \cdot {\bf P}$ is called
flow alignment, and imitates the shape of
a filament such as rod-like ($\xi>0$) or disk-like $\xi <0$.
It also demonstrates 
shear-alignment ($|\xi|>1$) or shear-tumbling ($|\xi|<1$) depending on the
parameter $\xi$.
Here, $v_p$ is the polymerization velocity and describes advection.
We neglect this term here and set $v_p=0$ because this is not the main
mechanism of self-propulsion in this model.

As we are interested in small systems with the characteristic length
scale $l \simeq 0.1-10 \mu$ m, we use the Stokes equation:
\begin{align}
\eta \nabla^2 {\bf v} - \nabla p
+ {\bf f} 
 &=0
 ,
 \label{Stokes}
\end{align}
where the force acting on the fluid is
\begin{align}
{\bf f}
&=
{\rm div} \left(
\sigma^{(a)} + \sigma^{(e)}
\right).
\end{align}
The force acting on the fluid has two parts: active stress and elastic stress.
The active stress is given by  \cite{kruse:2005,Tjhung:2012}
\begin{align}
\sigma_{ij}^{(a)} 
&=
\zeta P_i P_j
.
\end{align}
The sign of the parameter $\zeta$ demonstrates contractile stress
($\zeta>0$) and extensile stress ($\zeta<0$). 
The elastic stress $\sigma^(e)$ arises from the Frank elasticity of the
polar field and under a constant approximation
\begin{align}
\sigma_{ij}^{(e)} 
 &=
 - \frac{\xi}{2} \left(
P_i h_j + P_j h_i
 \right)
-K \nabla_i P_k \nabla_j P_k
,
\end{align}
where $\bf h = - \delta F/ \delta {\bf p}$.

As we are interested in the universal aspects of the model, we simplify
it so that it can reproduce spontaneous motion.
It is often the case that friction of the velocity field is included in
(\ref{Stokes}) to express friction between a cell and
substrate.
This term is given by $- \kappa {\bf v}$, which introduces another
length scale $\sqrt{\eta/\kappa}$ into the model.
The length scale sets screening of the hydrodynamic flow, and when the
friction is stronger, the velocity gradient is more localized near the
interface between the drop and surrounding fluid.
Although this term changes a critical activity to obtain self-propulsion,
it does not change the structure of bifurcation and therefore it is
not the main mechanism of motility in this model.
We, thus, neglect this term to consider the minimal ingredients of self-propulsion.

The free energy is chosen as (\ref{FreeEnergy.p}) so that the polar state $|{\bf P}=1|$ is stable
and thus the equation of the polarity field is given by
\begin{align}
 \pdiff{{\bf P}}{t}
+ \omega \cdot {\bf P}
&=
\xi {\bf \kappa} \cdot {\bf P}
 + \frac{1}{\Gamma}
 {\bf P} \left(
1 - |{\bf P}|^2
\right)
+ \frac{K}{\Gamma} \Delta {\bf P}
.
\end{align}
Together with (\ref{phase.field}), we have the closed equations.

\subsection{Numerical Simulation}

To consider translational motion and deformation of the drop, the
density field $\phi({\bf x})$ is introduced using the phase-field approach
\begin{align}
&
\pdiff{\phi}{t}
+ {\bf v} \cdot \nabla \phi 
\nonumber \\
 = &
D \nabla^2 \phi
+ g \phi(1 - \phi)
\left( \phi - \frac{1}{2} + \alpha \delta - \frac{6\beta}{g} (1-|{\bf p}|^2)
 \right)
 ,
\label{phase.field}
 \end{align}
where
\begin{align}
\delta 
&=
V_0 -
\int 
\phi^2 
(3 - 2 \phi )
 dV
 .
\end{align}
The equation has the two stable points $\phi=0$ and $\phi=1$, and they
indicate inside ($\phi=1$) and outside ($\phi=0$) the drop.
Although the conventional Ginzburg-Landau (GL) equation has an infinitely growing domain, the
modified version suppresses the growth to describe a drop with a finite size.
When the volume is equal to $V_0$, $\phi=1/2$ is an unstable point as in
the
standard time-dependent Ginzburg-Landau (TDGL) model.
The difference is found in $\delta (\phi)$, which controls the unstable point
so that the total volume is approximately conserved as $V_0$.
The advantage of this approach is that the model has a free energy \cite{Nonomura:2012}
\begin{align}
F_{\rm GL}
=&
\int \left[
\frac{g}{4}
\phi^2 (1-\phi^2)
 + \frac{\epsilon^2}{2} |\nabla \phi|^2
\right] dV
\nonumber \\
&+ \frac{\alpha g}{12}
\left(
V_0 - \int \phi^2 (3-2\phi) dV'
\right)^2.
\end{align}
The additional term makes a contribution to the hydrodynamic equation as
a gradient term, and this does not modify the form of the force due to
the phase field.

The free energy of this model is given by the following three terms
\begin{align}
F 
&=
 F_{\rm GL}
+ F_{\rm p}
+ F_{\rm cp}
\\
 F_{\rm p}[{\bf p}({\bf x})] 
&=
\int dV \left[
 \left(
\frac{1}{2} |{\bf p}|^2
+ \frac{1}{4} |{\bf p}|^4
\right)
+ \frac{K}{2} |\nabla_i p_j|^2
 \right]
 \label{FreeEnergy.p}
\\
 F_{\rm cp}[\phi({\bf x}),{\bf p}({\bf x})] 
&=
\beta
\int dV \left[
(1-| {\bf p} |^2) 
\phi^2
\left(
3 - 2 \phi
\right)
\right]
.
\end{align}
The free energy of the polarity field ensures that the amplitude of the
polarity is $|{\bf P}|=1$ almost everywhere except in the region in which
the polarity field is largely deformed.
The deformation of the polarity field is penalized by the Frank elastic
constant, $K$.
The indices in the elastic term assumes a matrix norm.
The coupling between $\phi$ and ${\bf p}$ is chosen so that $|{\bf
p}|=1$ inside the drop where $\phi=1$ and $|{\bf p}|=0$ outside.
For this purpose, we set $\beta = 1$.


The velocity field is calculated in the Fourier space.
The velocity field is expressed as
\begin{align}
\tilde{{\bf v}}({\bf k}) 
&=
\tilde{T}({\bf k}) 
\cdot \tilde{{\bf f}}({\bf k}),
\end{align}
where the Oseen tensor in the Fourier space is 
\begin{align}
\tilde{T}_{ij} (\bf k) 
&=
\frac{1}{\eta k^2}
\left(
\delta_{ij} - \frac{k_i k_j}{k^2}
\right).
\end{align}



\section{Spontaneous Motion and Deformation}

\begin{figure}[h]
 \begin{center}
  \includegraphics[width=0.50\textwidth]{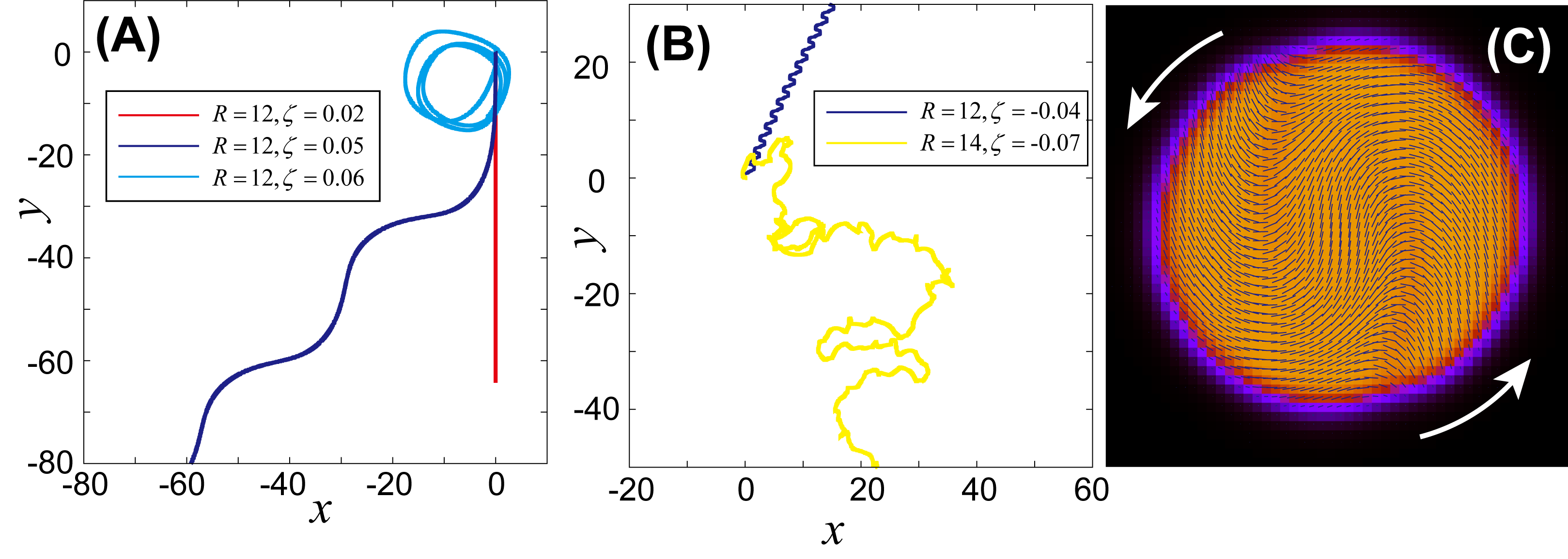}
  \caption{ (Color Online)
  Typical trajectories of active polar drops in different types of
  motion with different activities $\zeta$. (A) Contractile stress for
  $R=12$ with $\zeta=0.02$ (translation, red), $\zeta=0.05$ (zigzag
  motion, blue) $\zeta=0.06$ (rotation, light blue). (B) Extensile
  stress $R=12, \zeta=-0.04$ (zigzag motion, blue) and $R=14,
  \zeta=-0.07$ (chaotic motion, yellow). (C) Polarity field ${\bf P}$
  and its amplitude $|{\bf P}|$ of spinning active drop with the extensile
  stress $R=12$ and $\zeta=-0.03$.
  The direction of rotation is shown by white arrows.
  The parameters are chosen as $\eta=1.0$, $K=0.04$, $\xi$=1.1, $g = 0.2$, $\beta=1.0$,
  $\alpha=1.0$, and $D=1.0$.
\label{fig.pusher}
}
 \end{center}
\end{figure}

First, we discuss our numerical results.
The active polar drops are stationary when their
activity is low and their size is small.
 As the activity increases, the stationary state becomes unstable and
 the drop starts to move.
 The motion is straight with a constant speed as shown in
 Fig.~\ref{fig.pusher}(A).
 This motion occurs both in extensile and contractile drops.
 This result has already been obtained in \cite{Tjhung:2012}.
 However, as the activity increases, the extensile drop produces
 different motion to the contractile drop.
 The extensile drop has a spinning motion where the center of mass does
 not move, but the polarity field changes by rotating the drop (Fig.~\ref{fig.pusher}(C)).
 Rotational motion also occurs in which the translational motion follows
 a closed path (Fig.~\ref{fig.pusher}(A)).
 Then, as the activity increases, the motion of the extensile drop
 changes from a zigzag motion to random motion (Fig.~\ref{fig.pusher}(B)).
 These complex motions do not occur in the contractile drop.
 Such chaotic motion has not been obtained by the hydrodynamic model,
 but has been observed in extensile drops using
 the kinetic model\cite{Gao:2017}. 
 
\begin{figure}[h]
\begin{center}
\includegraphics[width=0.50\textwidth]{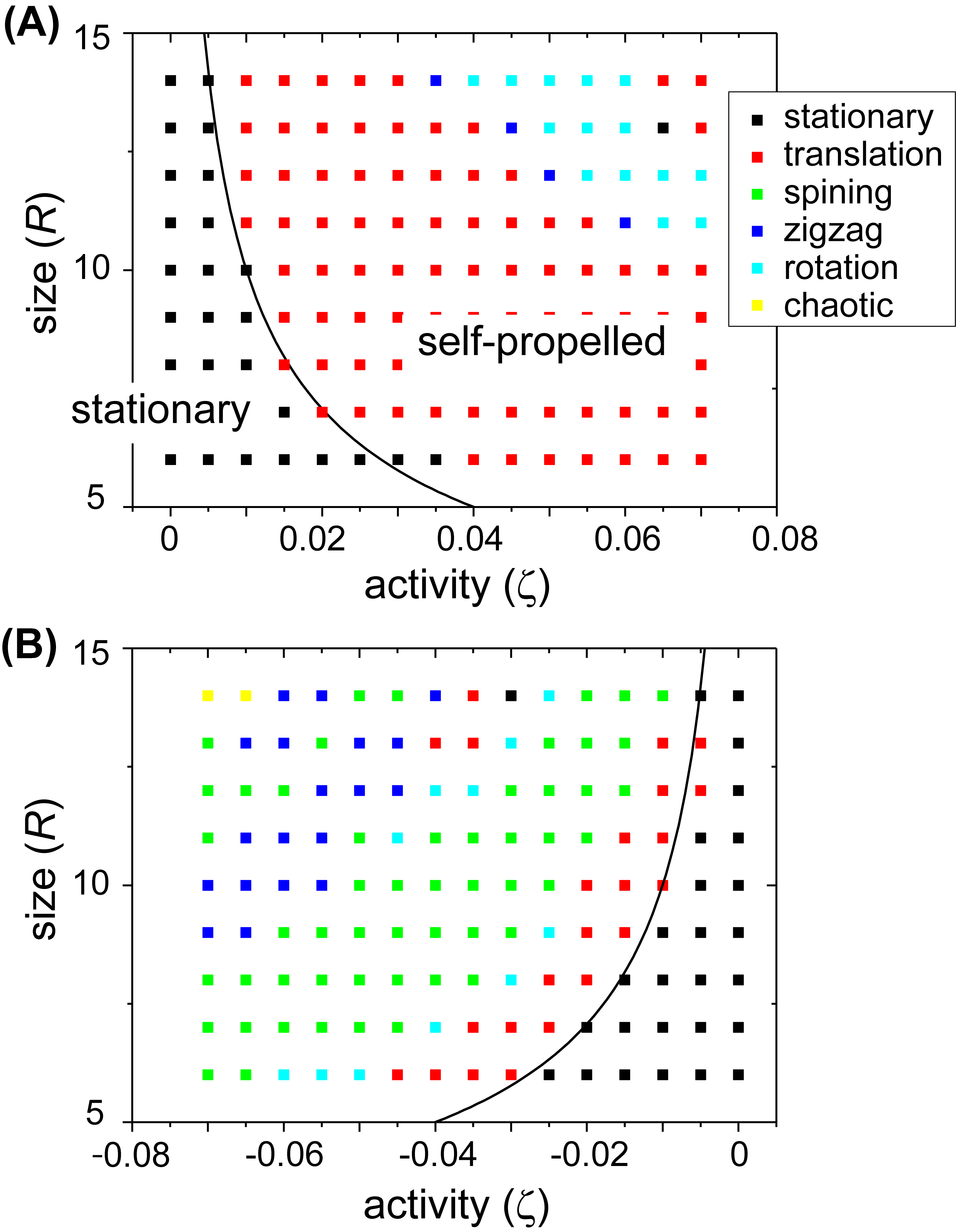}
\caption{ (Color Online) 
Phase diagrams of the motion of active polar drops with (A) contractile
 and (B) extensile stress under different activities $\zeta$ and sizes
 $R$.
 The solid lines indicate $R=1/\sqrt{|\zeta|}$.
 \label{fig.phasediagram}
}
\end{center}
\end{figure}

The different behaviors of spontaneous motion between the contractile
and extensile drops are evident in the phase diagram shown in
Fig.~\ref{fig.phasediagram}.
The overall tendency is that for higher activity and larger drops, the self-propulsive motion becomes complex.
The transition between the stationary state and the straight translational
motion is characterized by a non-dimensional number of the P\'{e}clet number,
Pe,
\begin{align}
  {\rm Pe}
 &=
 \frac{U}{R} \frac{R^2 \Gamma}{K}
 =
 \frac{|\zeta| R^2 \Gamma}{\eta K}
 .
 \label{eq.Peclet}
\end{align}
In fact, the transition between stationary and self-propelled states
occurs at ${\rm Pe} \simeq 25$ as shown in Fig.~\ref{fig.phasediagram}.
At higher activity, the boundary between two types of the motion becomes
less clearer.
In the phase diagram, we define each state as {\it stationary} for
$\langle |u| \rangle \leq 0.001$ and $\langle |\omega| \rangle \leq 0.01$,
{\it translation} for $\langle |u| \rangle > 0.001$
and $\langle |\omega| \rangle \leq 0.01$,
{\it spinning} for $\langle |u| \rangle \leq 0.001$,
$\langle |\omega| \rangle > 0.01$.
When $\langle |u| \rangle > 0.001$ and $\langle |\omega| \rangle > 0.01$,
the motion is called
{\it zigzag} for
$({\rm max} \omega - {\rm min} \omega)/{\rm max} |\omega| \geq 1$
and
{\it rotation} for
$({\rm max} \omega - {\rm min} \omega)/{\rm max} |\omega| < 1$.
This criteria ensures whether the drop periodically turns left and right
(zigzag), or moves with constant angular velocity (rotation).
When the speed and angular velocity changes irregularly, its state is called {\it chaotic}.

To summarize the numerical results, several observation can be made.
(i) Spontaneous motion occurs irrespective of the sign of active
stress.
The contractile and extensile drops both exhibit self-propulsion, and
seem to be unstable above the critical activity $\zeta_c$.
For an extensile drop, this is not surprising as the bend instability
should occur in unconfined systems for $\xi >1$.
However, the contractile active polar fluid in large systems is linearly
stable\cite{Edwards:2009}, while both our numerical results and those in \cite{Tjhung:2012,Marth:2015} indicate the pitchfork bifurcation.
(ii) Both the hydrodynamic model\cite{Tjhung:2012,Marth:2015} studied here using partial differential
equations of density and polarity fields and the kinetic model \cite{Gao:2017} using a
Smoluchowski equation of a probability distribution of a position and
orientation of filaments reproduce self-propulsion including chaotic
motion.
This implies that the mechanism of various types of self-propulsion is
encoded in the hydrodynamic model, while to have quantitative
features, one has to consider the kinetic model, which includes more
information about higher-order moments.

\section{Theoretical Analysis}

To study the mechanism of self-propulsion, we consider the polarity field inside a disk with a radius $R$.
When there is no disclination, the polarity is a unit vector and thus
is expressed by
\begin{align}
 {\bf P}
 &=
 \left(
\cos \psi, \sin \psi
 \right)
 .
\end{align}
The phase $\psi({\bf x})$ may be expanded in polar coordinates as
\begin{align}
 \psi
 &=
 \sum_{m=0}^{\infty}
 \left[
 \psi_{m} (r) \cos m\theta
 + \tilde{\psi}_m (r) \sin m \theta
 \right]
 \label{polar.phase.expansion}
 .
\end{align}
The advantage of this approach is that the bend (splay) instability
is expressed by $m=1$ and $\psi = A r$ ($\tilde{\psi}=Ar$).
It is also convenient to expand other vector fields such as velocity
${\bf v}$ and force ${\bf f}$ fields\cite{Yoshinaga:2018c}, for example,
\begin{align}
 {\bf v}
 &=
 \sum_m
 \left[
 \left(
{\bf v}_{r,m} \cdot {\bf r}_m
 \right) {\bf n}
 +
  \left(
{\bf v}_{t,m} \cdot {\bf t}_m
 \right) {\bf t}
 \right]
 ,
 \label{vel.expansion}
\end{align}
where ${\bf n}$ and ${\bf t}$ are unit normal and tangential vectors,
respectively (see Fig.~\ref{fig.isolated.flow.dipole}), and we define
two unit vectors as $
 {\bf r}_m
 \equiv
 \left(
\cos m \theta, \sin m \theta
 \right)
 $ and $
 {\bf t}_m
 \equiv
 \left(
-\sin m \theta, \cos m \theta
 \right)
 $.
For the special case,  ${\bf r}_1 = {\bf n}$ and ${\bf t}_1 = {\bf t}$.

\begin{figure}[h]
\begin{center}
 \includegraphics[width=0.50\textwidth]{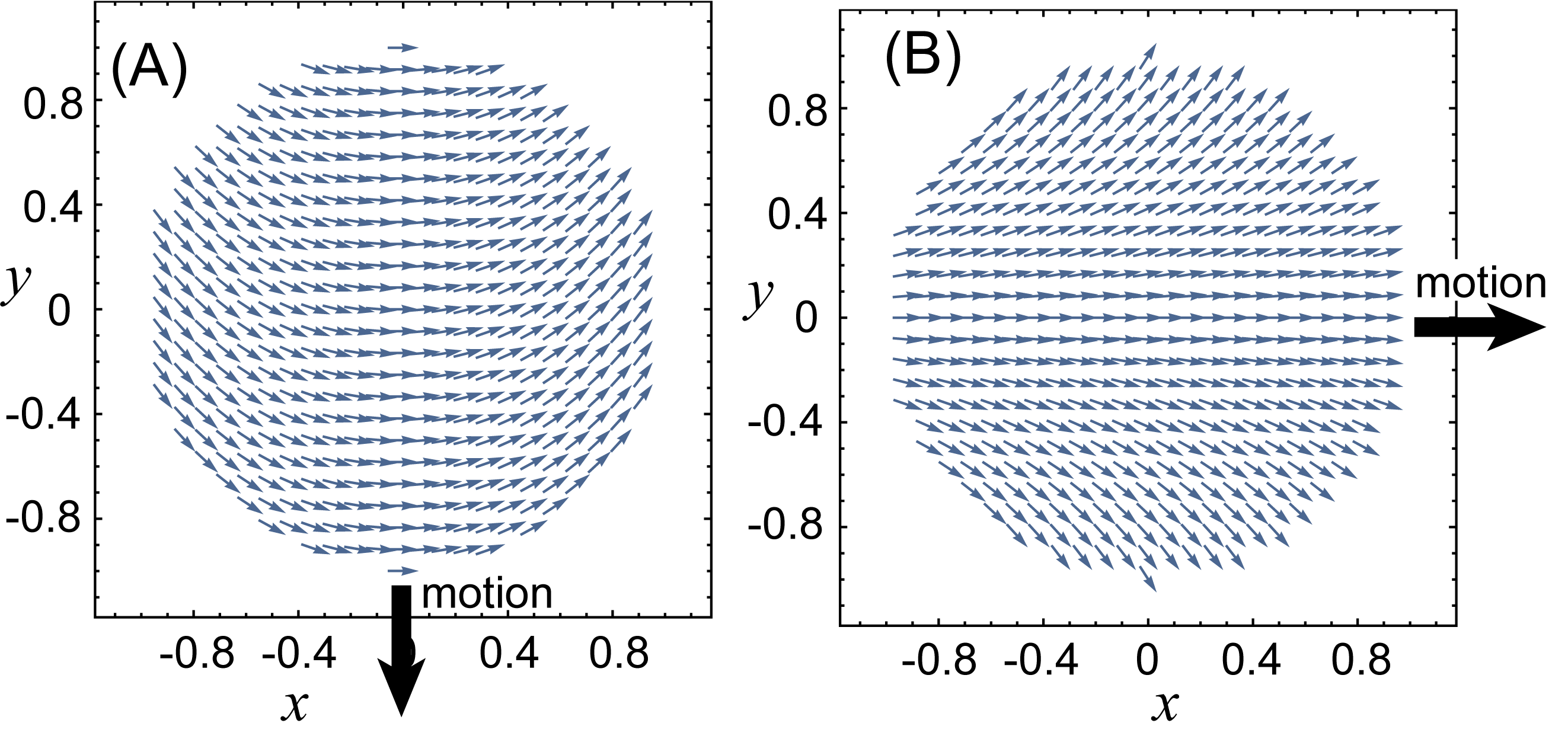}
\caption{ (Color Online) The polarity fields expressed by (\ref{polar.phase.expansion})
 where non-zero coefficients are only (A) $\psi_1 = 0.4 r$ and (B)
 $\tilde{\psi}_1 = 0.4 r$.
 The directions of motion are shown by the black arrows for (A)
 extensile and (B) contractile stress.
\label{fig.psi.expansion}
}
\end{center}
\end{figure}

By projecting (\ref{active.polar2.drop.model}) onto $(-\sin \psi, \cos
\psi)$, we obtain the equation of the phase as
\begin{align}
&
 \partial_t \psi
 =
 \omega_{xy}
 - \left(
v_x \partial_x + v_y \partial_y
 \right) \psi
 \nonumber \\
 &
 + \xi \left[
 \frac{\kappa_{yy} - \kappa_{xx}}{2} \sin 2 \psi
 + \kappa_{xy} \cos 2 \psi
 \right]
 + \Gamma^{-1} K  \Delta \psi
 .
\end{align}


We choose the unit length scale as $R$, the unit time scale as
$R/U_0$, the characteristic velocity as $U_0 = \frac{\zeta R}{\eta}$, and the
characteristic pressure as $p_0=\zeta$. 
The non-dimensional numbers in this system are the Peclet number (\ref{eq.Peclet}) and the
Ericksen number
\begin{align}
 {\rm Er}
 &=
 \frac{K}{|\zeta| R^2}
 .
\end{align}
The non-dimensionalized equations are obtained as
  \begin{align}
&   
  \Delta \psi
  =
    {\rm Pe} \left(
  \partial_t \psi
- \omega_{xy}
   +
{\bf v} \cdot \nabla
   \psi
   \right.
   \nonumber \\
&
   \left.
 - \xi \left[
 \frac{\kappa_{yy} - \kappa_{xx}}{2} \sin 2 \psi
 + \kappa_{xy} \cos 2 \psi
 \right]
  \right)
   \label{active.polar2.drop.model.normalise} 
  \\
&
   \Delta {\bf v} - \nabla p + {\bf f}
   =0
     \label{active.polar2.drop.model.normalise.stokes} 
   \\
   &
  {\rm div} {\bf v}
  =
   0
  \label{active.polar2.drop.model.normalise.divv} 
   \\
   &
  {\bf f}
  =
  {\rm div} \left(
\pm
  {\bf P} {\bf P}
  +
   {\rm Er}
   \sigma^{(e)}
   \right)
 .
  \end{align}
Here, the sign in the active stress is chosen such that positive ($\zeta>0$)
corresponds to contractile stress whereas negative ($\zeta<0$) corresponds to
extensile stress.
Without topological defects, the amplitude of the polarity field is
almost constant $|{\bf P}| \simeq 1$ and therefore the elastic stress is
$\sigma^{(e)} \sim \left(  \partial {\bf P}  \right)\left(  \partial{\bf P}  \right) $.
This term is neglected in our analysis as it is not necessary to
reproduce self-propulsion.
We have confirmed that the numerical simulations without elastic
stress also demonstrate self-propulsion of the active polar drop.

Boundary conditions for (\ref{active.polar2.drop.model.normalise.stokes})-(\ref{active.polar2.drop.model.normalise.divv}) at $r=1$ are
\begin{align}
{\bf n} \cdot {\bf v}^{(i)}
 &=
 {\bf n} \cdot {\bf v}^{(o)}
 = {\bf u} \cdot{\bf n}
 \\
{\bf t} \cdot {\bf v}^{(i)}
 &=
 {\bf t} \cdot {\bf v}^{(o)}
 \\
 {\bf n} \cdot
 \bm{\sigma}^{(o)}
 &=
 {\bf n} \cdot
 \bm{\sigma}^{(i)}
 +
  {\bf n} \cdot
 \left(
\pm
  {\bf P} {\bf P}
  +
  {\rm Er}
 \sigma^{(e)}
 \right)
 \label{BC.stress}
 .
\end{align}
Here, ${\bf u}$ is self-propulsive velocity.
The first two equations demonstrate continuity of the velocity fields
inside and outside the drop.
The last condition implies force balance on the surface of the drop. 
There is another boundary condition for $\psi$ in
(\ref{active.polar2.drop.model.normalise}).
This is set by the condition that there is no mechanical force acting in
the system (force-free condition) in (\ref{psi.BC}).


\subsection{Self-propulsive Motion}

In this section, we consider how self-propulsion occurs in the active
polar drop.
The velocity of a drop is computed as\cite{Yabunaka:2012,Yoshinaga:2014}
\begin{align}
 {\bf u}
 &=
 \frac{1}{\pi R^2}
 \int v_n {\bf R} dS,
 \label{velocity.geometric}
\end{align}
where ${\bf R}$ is a vector pointing to the surface of the drop.
The normal velocity on the surface is
$
 v_n
 =
 {\bf v} \cdot {\bf n}
$.
For a circular drop, the velocity given by (\ref{velocity.geometric}) is
non-zero only when $v_n \sim \cos \theta$ or $v_n \sim \sin \theta$.
We expand the velocity field using force multipoles, $F_{i_1 i_2 \cdots i_l}$,  as
\begin{align}
 v_i
 &=
 T_{ij} F_j
 - \partial_k T_{ij} F_{kj}
 +  \partial_k \partial_l T_{ij} F_{klj}
 - \cdots,
\end{align}
where $T_{ij}({\bf x},{\bf x}')$ is the Oseen tensor, and the $l$th force
multipoles are expressed as
\begin{align}
 F_{i_1 i_2 \cdots i_l}
 &=
 \int
 x_{i_1} x_{i_2} \cdots x_{i_{l-1}}
 f_{l}
 .
 \label{force.mulitpole}
\end{align}
Under the force-free condition, the self-propulsive velocity (\ref{velocity.geometric}) is nonzero
only when there is a source dipole.
The velocity field of the source dipole in two dimensions is obtained by
taking trace ($\delta_{kl}$) of $\partial_k \partial_l T_{ij}$ as
\begin{align}
 {\bf v}^{(q)}
 &=
 \frac{1}{2\pi \eta r^2}
 \left[
  ({\bf q} \cdot {\bf r}) {\bf r}
 -
 ({\bf q} \cdot {\bf t}) {\bf t}
 \right]
 ,
\end{align}
where the source dipole is obtained from the distribution of the force as
\begin{align}
 {\bf q}
 &=
 \frac{1}{8}
 \int
 \left(
 3 r^2 {\bf f}
 - 2 {\bf x} ({\bf x} \cdot {\bf f})
 \right) dV
 \label{source.dipole}
 .
\end{align}
For a given source dipole, the velocity is obtained as
\begin{align}
 {\bf u}
 &=
 \frac{1}{2 \pi \eta R^2}
 {\bf q}
 ,
\end{align}

This aspect of the self-propulsion is shared by other phenomena such as
squirmers\cite{Lauga:2009}, phoretic motion and self-phoresis\cite{julicher2009}, and self-propulsion driven by
the Marangoni effect\cite{Yoshinaga:2014}.
All these motions are driven by the source dipole of their flow fields.
In contrast to the active polar drop, the flow field in these models
is driven by interfacial (or surface) force\cite{anderson:1989}.
In the active drop studied here, the flow is driven  by active stress
acting in bulk.
Nevertheless, in terms of force-free motion, only the source dipole is
associated with self-propulsion.

The active stress is expressed in terms of polynomials of $\psi$
 \begin{align}
  \sigma^{(a)} - {\rm Tr} \sigma^{(a)}
  &=
  \pm
\begin{pmatrix}
 \cos 2\psi & \sin 2 \psi\\
 \sin 2\psi & -\cos 2 \psi
\end{pmatrix}  
  \nonumber \\
  &=
  \pm
  \left[
\begin{pmatrix}
 1 & 0\\
 0 & -1
\end{pmatrix}
  +
\begin{pmatrix}
 0 & 2\psi\\
 2 \psi & 0
\end{pmatrix}
  +
\mathcal{O} (\psi^2)
  \right]
  .
 \end{align}
 We subtract a trace of the active stress as it merely modifies the pressure.
By expansion of (\ref{polar.phase.expansion}), the self-propulsive
velocity is also expanded in terms of its coefficients $\psi_m$ and
$\tilde{\psi}_m$ as
\begin{align}
 {\bf u}
 &=
 \pm \frac{1}{\eta R^2}
 \left[
 \int dr r^2 \odiff{}{r} r
 \begin{pmatrix}
\tilde{\psi}_1 \\
 \psi_1  
 \end{pmatrix}
 + \mathcal{O} \left(
 \psi_m^2, \psi_m \tilde{\psi}_m, \tilde{\psi}_m^2
 \right)
 \right]
 .
\end{align}
When $\psi(r) = a_1^1 r$ and $\tilde{\psi}(r) = b_1^1 r$ with a given
constants $a_1^1,b_1^1$, the velocity
is ${\bf u} \sim \pm(b_1^1,a_1^1)$.
This analysis suggests that, for small distortion of the polarity field,
the velocity of the drop is proportional to the first mode of the
expansion in (\ref{polar.phase.expansion}).
The contractile active stress gives rise to self-propulsion in the $+x$
direction for $\tilde{\psi}_1 > 0$, whereas the extensile active stress
results in motion in the $-y$ direction for $\psi_1 > 0$ (see Fig.~\ref{fig.psi.expansion}). 

\subsection{Perturbation around Stationary States}

When ${\rm Pe}=0$, the stationary polarity field is uniform inside the
drop.
As ${\rm Pe}$ increases, the flow field is generated by the active
stress.
When ${\rm Pe}$ is small, the flow may perturb the polarity field, but
it is not strong enough to generate a source dipole.
The flow is dipolar and does not lead to
self-propulsion.
The drop remains stationary.
To study the mechanism of self-propulsion, we need to clarify when this
stationary state becomes unstable.
Therefore, we linearize the model around the stationary state.

In contrast to self-propulsion driven by chemical reactions through
the Marangoni effect \cite{Yabunaka:2012,Yoshinaga:2014} in which the
stationary state is trivial, the stationary state of the active polar
drop is not simple owing to the flow alignment term.
The uniform polarity field does not satisfy
(\ref{active.polar2.drop.model.normalise}) owing to the stress acting on
the surface of the drop.
Nevertheless, it is important to observe that
(\ref{polar.phase.expansion}) is divided into two terms: odd and even
$m$.
The stationary state results in dipolar flow, which generates perturbation
only in even $m$ terms. 
This implies that when the initial phase does not contain odd modes, there
is no odd mode in the final state.
On the other hand, self-propulsion is associated with the first mode of (\ref{polar.phase.expansion}).
As we are interested in transition between the stationary state and
self-propelled states, it is natural to use a following scaling
\begin{align}
 \psi
 &\sim
 \begin{cases}
  \mathcal{O}(1)
  \mbox{   for   }
  \mbox{even } m
  \\
  \epsilon
  \mbox{   for   }
  m=1
  \\
  \epsilon^2
  \mbox{   for   }
  \mbox{odd } m \neq 1
 \end{cases}
 \label{active.nematics.scaling}
\end{align}
We denote by $\psi^*$ a stationary solution of
(\ref{active.polar2.drop.model.normalise}), which contains only even
modes in (\ref{polar.phase.expansion}).
We linearize a set of the model (\ref{active.polar2.drop.model.normalise}) in terms of $\epsilon$.
Each variable is expanded as
\begin{align}
 \psi
 &=
 \psi^*
 + \epsilon \psi^{(1)} +
 \epsilon^2 \psi^{(2)} + \cdots
 \label{active.polar.drop.psi.epsilon.expansion}
\end{align}
and following (\ref{polar.phase.expansion}), each order in
(\ref{active.polar.drop.psi.epsilon.expansion}) is expanded in modes
denoted by $m$.
Note that $\psi^*$ contains only even $m$ whereas $\psi^{(1)}$ contains
$m=1$ and all even $m$.
Higher-order terms such as $\psi^{(2)}$ contain, in general, all $m$.
The active stress is expanded accordingly as
 \begin{align}
  \sigma^{(a)} - {\rm Tr} \sigma^{(a)}
    & =
  \sigma^{(a,*)} +   \epsilon \sigma^{(a,1)} \psi^{(1)}
  + \cdots
  \\
  \sigma^{(a,*)}
  &=
  \pm
    \begin{pmatrix}
 \cos 2\psi^* & \sin 2 \psi^* \\
 \sin 2\psi^* & -\cos 2 \psi^*
  \end{pmatrix}
  \\
  \sigma^{(a,1)}
  &=
\pm 2  
  \begin{pmatrix}
 -\sin 2\psi^* & \cos 2 \psi^* \\
 \cos 2\psi & \sin 2 \psi^*
  \end{pmatrix}
  \label{active.stress.a1}
  .
\end{align}

At $\mathcal{O}(\epsilon)$, the model becomes,
\begin{align}
 \Delta \psi^{(1)}
 =&
 {\rm Pe}
 \left(
 - \omega_{xy}^{(1)}
 +
{\bf v}^* \cdot \nabla
 \psi^{(1)}
 +
{\bf v}^{(1)} \cdot \nabla
 \psi^*
  \right.
 \nonumber \\
 &
  \left.
 - \xi
 \left[
 \frac{\kappa_{yy}^* - \kappa_{xx}^*}{2}
  \cos 2 \psi^*
 + \kappa_{xy}^*   \sin 2 \psi^*
 \right]\psi^{(1)}
 \right.
 \nonumber \\
 &
  \left.
 - \xi
 \left[
 \frac{\kappa_{yy}^{(1)} - \kappa_{xx}^{(1)}}{2}
  \sin 2 \psi^*
 + \kappa_{xy} ^{(1)}  \cos 2 \psi^*
 \right]
 \right),
 \label{active.nematics.linear.psi}
\end{align}
where the velocity gradient and rotation tensors are proportional to
$\psi^{(1)}$ as
$ \kappa_{ij}^{(1)}
 =
 \frac{1}{2} \left(
 \partial_i v_j^{(1)} + \partial_j v_i^{(1)}
 \right)
 $
and 
$  \omega_{ij}^{(1)}
 =
 \frac{1}{2} \left(
 \partial_i v_j^{(1)} + \partial_j v_i^{(1)}
 \right)
 $.
 The velocity field perturbed by $\psi^{(1)}$ is the solution of
 (\ref{active.polar2.drop.model.normalise.stokes}) under the active
 stress (\ref{active.stress.a1}).
We are interested in $\psi^{(1)}_1$: when it becomes non-zero as ${\rm
Pe}$ is varied and how it grows.
Thanks to the scaling (\ref{active.nematics.scaling}), the even modes in
$\psi^{(1)}$ do not generate $\psi^{(1)}_1$ because $\psi^*$ consists
only of even modes.
Therefore, (\ref{active.nematics.linear.psi}) is formally rewritten as
\begin{align}
 \Delta \psi^{(1)}_1
 &=
 {\rm Pe}
 \mathcal{N}^{(1)} \psi^{(1)}_1
 .
 \label{linearise.eq}
\end{align}
We analyze (\ref{linearise.eq}) by using the Zernike expansion of $\psi({\bf x})$ is given by
\begin{align}
 \psi(r,\theta)
 &=
 \sum_{n,m}
 \left[
 a^m_n
 R^m_n (r)
 \cos m\theta
 +
 b^m_n
 R^m_n(r)
 \sin m\theta 
 \right],
 \label{psi.Zernike}
\end{align}
where
\begin{align}
 R_n^m (r)
 &=
 \sum_{k=0}^{\frac{n-m}{2}}
 \frac{(-1)^k (n-k)!}{k! (\frac{n+m}{2}-k)! (\frac{n-m}{2}-k)!}
 r^{n-2k}
 \label{Zernike.coefficients}
\end{align}
for even $n-m$. For odd $n-m$, then $R_n^m=0$.
Note that $n \geq m$.
Clearly, for given $n \in [0,\infty)$ and $n \geq m$, $R_n^m(r)$ is
expressed by power series in $r$.

Taking inner product of the linearized
equation (\ref{linearise.eq}) with the Zernike polynomials for $m=1$ in
the expansion of (\ref{psi.Zernike}), we obtain the following linear
algebraic equation using the boundary condition (\ref{psi.BC}):
\begin{align}
 L
 \begin{pmatrix}
  a_n^1 \\
  b_n^1
 \end{pmatrix}
 &\equiv
   \begin{pmatrix}
  L^{(aa)} & L^{(ab)} \\
    L^{(ba)} & L^{(bb)} 
 \end{pmatrix}
 \begin{pmatrix}
  a_n^1 \\
  b_n^1
 \end{pmatrix}
 =0
 ,
 \label{linearised.eq.matrix}
\end{align}
where each block is a square matrix and is given by
\begin{align}
 L^{(aa)}
 &=
 \begin{pmatrix}
  1 &   1 &   1 & \cdots \\
  0 & 6 & 16 & \cdots \\
  0 & 0 & 10 & \cdots \\
  \vdots & \vdots & \vdots & \ddots
 \end{pmatrix}
 - {\rm Pe} \delta L^{(aa)}
 \label{linear.matrix.aa}
\end{align}
and
\begin{align}
 L^{(ab)}
 &=
 {\bf 0} - {\rm Pe} \delta L^{(ab)}
 .
\end{align}
The other two matrices are expressed in a similar way.
The first term in (\ref{linear.matrix.aa}) corresponds to the left-hand
side in (\ref{linearise.eq}) owing to the Frank elasticity (see (\ref{Zernike.Laplace})), whereas the second term corresponds to the
right-hand side from the coupling with fluid flow.
The first row in the matrices represents the boundary condition
(\ref{psi.BC}), namely,
\begin{align}
 \sum_{n=1} a_n^1 R_n^{1}(1)
 &=
  \sum_{n=1} b_n^1 R_n^{1}(1)
 =0
 .
\end{align}
Other rows in the matrices are obtained by taking an inner product of
(\ref{linearise.eq}) for each basis in (\ref{psi.Zernike}), namely,
multiplying $R_n^1(r)$ and integrating over $r$ using the orthogonal
relation (\ref{Zernike.orthogonality}).
From the second row, the projection onto $R_1^1(r)$, $R_3^1(r), \ldots$
is performed.
When ${\rm Pe}=0$, the matrix (\ref{linear.matrix.aa}) is clearly
invertible and therefore the solution is $a_n^1=b_n^1=0$ and the
stationary solution is stable.
The diagonal term in (\ref{linear.matrix.aa}) is given by $2(k+2)$
for $k=1,3,\ldots$ corresponding to the second, third, $\ldots$ row.
This suggest that the possible scenario for zero eigenvalue arises from
the smallest $n$.

The stability of the stationary state $\psi^*$ is obtained by the
determinant of the linearized matrix in (\ref{linearised.eq.matrix}).
A non-trivial solution appears only when the determinant vanishes, and
the matrix is not invertible.
This corresponds to find zero eigenvalues in which the stationary state
looses it stability.
This aspect is explicitly shown in Appendix~\ref{sec.eigenvalues}.

\subsection{Stability around Uniform Orientation}

\begin{figure}[h]
\begin{center}
 \includegraphics[width=0.50\textwidth]{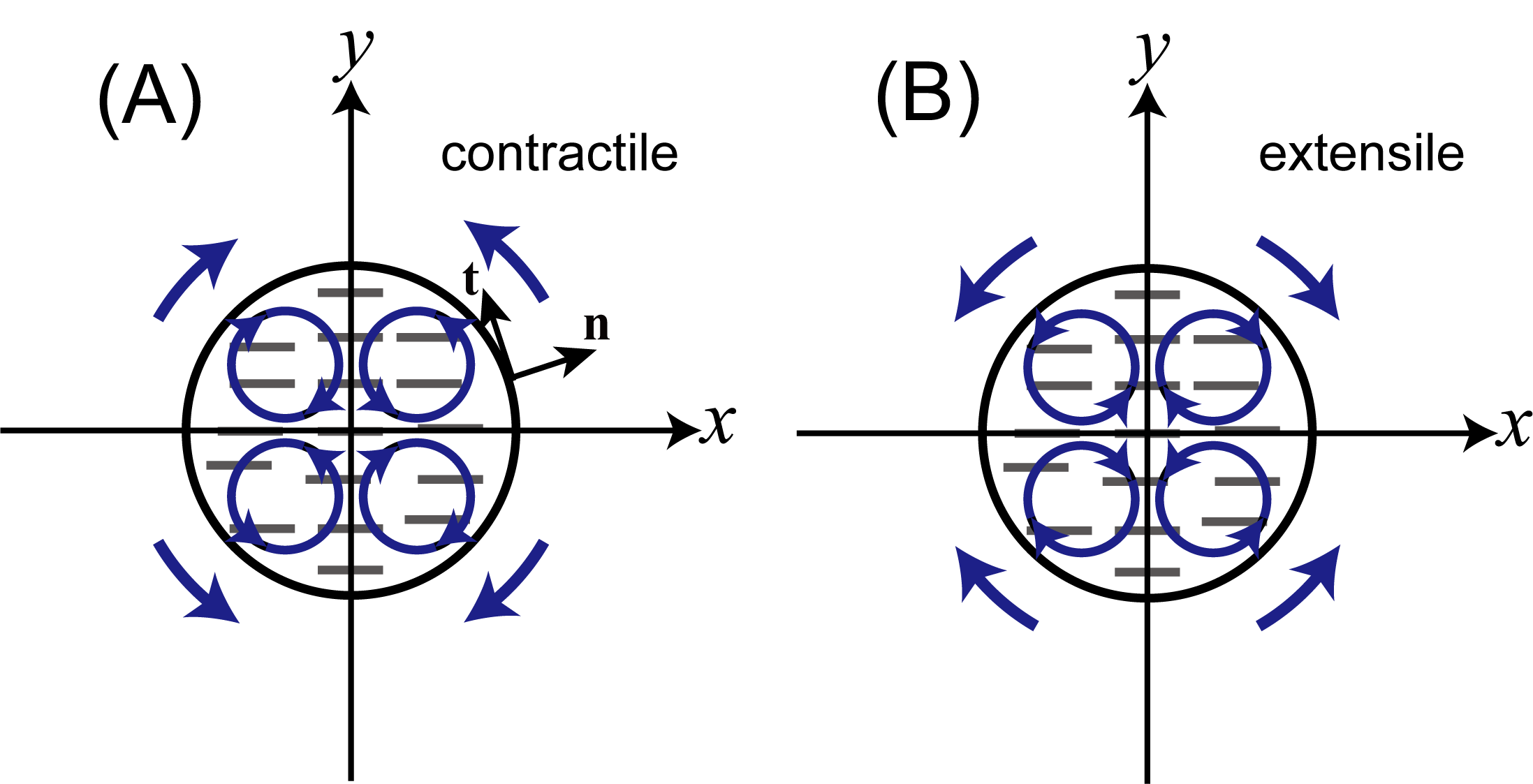}
 \caption{
 (Color Online) The steady velocity field ${\bf v}^*$ generated by the uniform polarity
 field $\psi^*$ for (A) contractile and (B) extensile active stress.
 The polarity field is shown by the gray line and the flow field is
 shown by blue arrows.
 The directions of a normal and tangential vectors are also shown by
 black arrows.
\label{fig.isolated.flow.dipole}
}
\end{center}
\end{figure}

To analyze (\ref{linearise.eq}), or equivalently
(\ref{linearised.eq.matrix}), we need to calculate stationary solution
of (\ref{active.polar2.drop.model.normalise})-(\ref{active.polar2.drop.model.normalise.divv}).
The concrete form of $\psi^*$ is available only numerically.
Nevertheless, it is reasonable to use an approximation
\begin{align}
 \psi^*
 &=
 0 + \mathcal{O}({\rm Pe})
 ,
 \label{simple.eq.psi}
\end{align}
where the polarity field is aligned with the $x$-axis (see Fig.~\ref{fig.isolated.flow.dipole}).
Here, the lowest-order term is chosen to be zero, though any spatially
uniform constant is
possible by rotation.
Then, the inhomogeneous term is approximated as
\begin{align}
 \mathcal{N}^{(1)} \psi^{(1)}_1
 &=
 - \omega_{xy}^{(1)}
 +
{\bf v}^* \cdot \nabla 
 \psi^{(1)}_1
 - \xi
 \frac{\kappa_{yy}^* - \kappa_{xx}^*}{2}
\psi^{(1)}_1
 - \xi
 \kappa_{xy} ^{(1)}
 .
 \label{simple.eq.linear}
\end{align}
Note that when $\psi^*=\pi/2$ is chosen
\begin{align}
 - \xi
 \frac{\kappa_{yy}^* - \kappa_{xx}^*}{2}
 \psi^{(1)}_1
 \rightarrow
 + \xi
 \frac{\kappa_{yy}^* - \kappa_{xx}^*}{2}
\psi^{(1)}_1
\end{align}
and, as we show later, the sign of the dipolar flow becomes
opposite.
As a result, the stability of (\ref{simple.eq.linear}) does not change.
The physical meaning of (\ref{simple.eq.linear}) is decomposed into two
parts.
The first part consists of the
first and last terms, which describe the flow generated by distortion of the
polarity field.
The second part consists of the second and third terms.
Flow is
generated by the {\it stationary} polarity field (\ref{simple.eq.psi}),
which is coupled with the distortion of the polarity field and
stabilizes/destabilizes the polarity field.


First, we compute steady velocity ${\bf v}^*$ and resulting shear flow
$\kappa^*$.
The active stress for the polarity field (\ref{simple.eq.psi}) is
expressed as
\begin{align}
 \sigma^{(a,*)}
 &=
\pm \Theta (R - r)
 \hat{\bf x}  \hat{\bf x}
 ,
\end{align}
where $\Theta(x)$ is a step function $\Theta(x)=1$ for $x>0$ and
$\Theta(x)=0$ otherwise.
The positive and negative signs denote contractile and extensile active
stress, respectively.
The active force generated by the active stress is obtained as
 \begin{align}
  {\bf f}
  &=
  f_n {\bf n}
  + f_t {\bf t}
  \\
  f_n
  &=
  \mp
  \frac{1}{2}
  \delta (R-r)
  \left(
1 + \cos 2\theta
  \right)
  \\
  f_t
  &=
  \pm \frac{1}{2}
  \delta (R-r) \sin 2\theta
  .
 \end{align}


 The steady velocity field is given by
 \begin{align}
  {\bf v}^*
  &=
  \left(
{\bf v}_n^* \cdot {\bf r}_2
  \right) {\bf n}
  +
  \left(
{\bf v}_t^* \cdot {\bf t}_2
  \right) {\bf t}
  ,
  \label{steady.velocity.n2}
 \end{align}
 where
 \begin{align}
  {\bf v}_r^*
  &=
  \begin{pmatrix}
   (r^3 - r) v_{2}^*  \\
   0
  \end{pmatrix}
  \\
    {\bf v}_t^*
  &=
  \begin{pmatrix}
   -(2 r - 4 r^3) \tilde{v}_{t,2}^*  \\
   0
   \end{pmatrix}
  .
 \end{align}
 Owing to incompressibility and the boundary condition of tangential stress at the
 interface, we obtain
 \begin{align}
  v_{2}^* &= \tilde{v}_{2}^*
  = \mp \frac{1}{16}
  .
 \end{align}
 Note that the balance of normal stress at the interface is satisfied by
 deformation.
 We assume the uniform surface tension is large so that deformation is
 negligibly small and therefore (\ref{steady.velocity.n2})
 approximates the velocity field under deformation.
 Schematic pictures of the velocity field are given in
 Fig.~\ref{fig.isolated.flow.dipole}.
 The extensile active stress generates dipolar flow pulling from the
 sides of the drop and pushing from top and bottom.
 The contractile active stress generates the flow in the opposite
 direction.
 The flow of contractile active stress is qualitatively the same as was
 observed in \cite{Tjhung:2012}. 
We finally obtain
\begin{align}
&
 {\bf v}^* \cdot \nabla \psi^{(1)}_1
\nonumber \\
 & =
 \sum_{m}
 \left[
 \left(
 v_{r,2}^* (r^3-r) a_m^1 R_m^{1'}(r)
 +
 v_{t,2}^* (2r-4r^3) b_m^1 R_m^{1'}(r)
 \right) \cos \theta
 \right.
 \nonumber \\
&
 \left.
 +
 \left(
-v_{r,2}^* (r^3-r) b_m^1 R_m^{1'}(r)
 +
 v_{t,2}^* (2r-4r^3) a_m^1 R_m^{1'}(r)
 \right) \sin \theta
 \right]
 \nonumber \\
 & + \mathcal{O} \left(
 \cos 3\theta, \sin 3\theta
 \right)
\end{align}
and
\begin{align}
&
 \xi
 \frac{\kappa_{xx}^* - \kappa_{yy}^*}{2}
 \psi^{(1)}
 \nonumber \\
 &=
 \frac{\xi}{4}
 \sum_{m}
    \left(
  v_{r,x}^{*'} + v_{t,x}^{*'}
  +  \left(
\frac{v_{r,x}^*}{r} + \frac{v_{t,x}^*}{r}
  \right)
 \right)
 \nonumber \\
&
 \times
 \left(
 a_m^1 R_m^1 (r)
   \cos \theta
 +  b_m^1 R_m^1 (r)
   \sin \theta
 \right)
 + \mathcal{O} \left(
 \cos 3\theta, \sin 3\theta
 \right)
 \label{stationary.kappa}
\end{align}


\begin{figure}[h]
\begin{center}
 \includegraphics[width=0.35\textwidth]{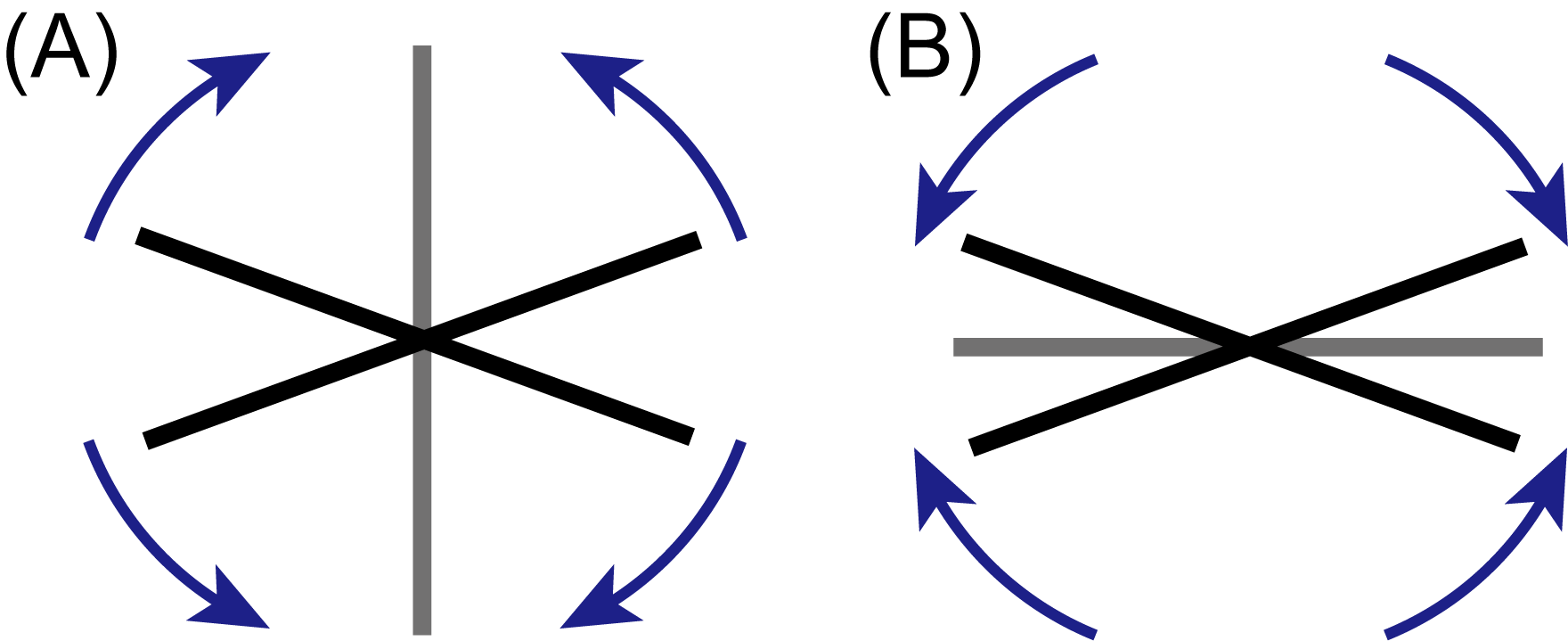}
 \caption{
(Color Online) The mechanism of rotation induced by dipolar flow generated by (A)
 contractile and (B) extensile active stress in stationary state.
 Initial orientation perturbed from the $x$-axis is shown in black
 lines, while stable states are shown in gray lines.
 \label{fig.isolated.shear.alignment}
}
\end{center}
\end{figure}

Next, we consider a velocity field generated by distortion of the polarity field $\psi^{(1)}_1$.
The active stress generated by $\psi^{(1)}_1$ is expressed as
\begin{align}
 \sigma^{(a,1)}
 &=
 \pm
 \begin{pmatrix}
  0 && 2 \\
  2 && 0
 \end{pmatrix}
 \psi^{(1)}
\end{align}
and the resulting force is
\begin{align}
 f^{(a,1)}
 &=
 \pm
 2 \bar{\nabla} \psi^{(1)}
\end{align}
with $\bar{\nabla}=(\nabla_y, \nabla_x)$.
The distortion is expressed by the first mode in (\ref{polar.phase.expansion})
which gives rise to the force
 .
\begin{align}
 {\bf f}^{(a,1)}
 &=
 ({\bf f}_{r}^{(a,1)} \cdot {\bf n}) {\bf n}
 +
 ({\bf f}_{t}^{(a,1)} \cdot {\bf t}) {\bf t}
 ,
\end{align}
where
\begin{align}
 {\bf f}_{r}^{(a,1)}
 &=
 {\bf f}_{t}^{(a,1)}
 =
 \pm
 \frac{1}{2}
 \begin{pmatrix}
  \frac{\tilde{\psi}^{(1)}_1}{r} + \tilde{\psi}^{(1)'}_1
  \\
    \frac{\psi^{(1)}_1}{r} + \psi^{(1)'}_1
 \end{pmatrix}
 =
 \pm
 \frac{1}{2 r}
 \odiff{}{r} r
  \begin{pmatrix}
  \tilde{\psi}^{(1)}_1
  \\
   \psi^{(1)}_1
 \end{pmatrix}
 .
 \label{force.expand.first.frft}
\end{align}
In order to ensure the force-free condition, it requires
 \begin{align}
  \psi^{(1)}_1(R) &= \tilde{\psi}^{(1)}_1=0
  .
 \label{psi.BC}
 \end{align}

We may express the velocity field at the first mode in
(\ref{vel.expansion}), and calculate it by the method outlines in Appendix~\ref{sec.vel}.
Once we obtain ${\bf v}_r^{(1)}$ and ${\bf v}_t^{(1)}$, the shear and
rotational flow is expressed as
\begin{align}
 \kappa_{xy}
 =&
 \frac{1}{4} \left(
v_{r,y}^{(1)'} + v_{t,y}^{(1)'}
 \right) \cos \theta
 +
  \frac{1}{4} \left(
v_{r,x}^{(1)'} + v_{t,x}^{(1)'}
 \right) \sin \theta
 \nonumber \\
 + &
 \mathcal{O}(\cos 3\theta, \sin 3 \theta)
 \label{kappa.m1}
  \\
 \omega_{xy}
 =&
- \frac{1}{2}
 \left(
 v_{n,x}^{(1)'} + 
v_{t,x}^{(1)'}
 \right) \sin \theta
 +
\frac{1}{2}
 \left(
 v_{n,y}^{(1)'} + 
v_{t,y}^{(1)'}
 \right) \cos \theta
 .
  \label{omega.m1}
\end{align}
Here, we have used incompressibility to simplify the rotational flow.
Using the force generated by active stress
(\ref{force.expand.first.frft}), the $\cos \theta$ terms in
(\ref{kappa.m1}) and (\ref{omega.m1}) result in $L^{(a,a)}$ in the linearized matrix
(\ref{linearised.eq.matrix}) for the coefficients of the expansion
(\ref{psi.Zernike}), whereas $\sin \theta$ terms result in $L^{(b,b)}$. 

\begin{figure}[h]
\begin{center}
\includegraphics[width=0.4\textwidth]{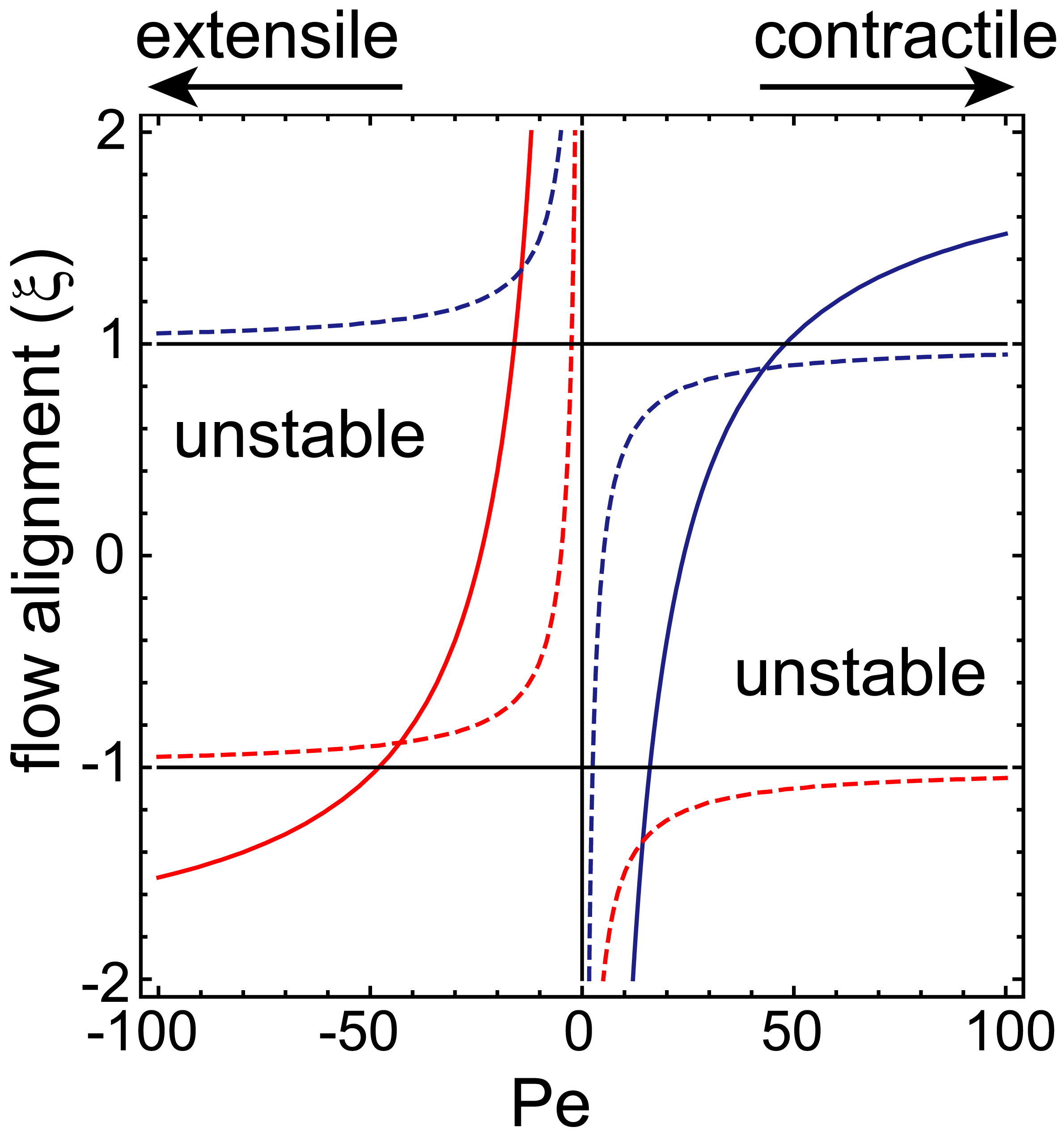}
 \caption{ (Color Online)
 Stability of the stationary state in which the polarity field is
 uniform inside the active polar drop.
 The stationary state is unstable in the region that does not include
 ${\rm Pe}=0$ for each line.
 The horizontal axis indicates signed Pe where ${\rm Pe}>0$
 demonstrates contractile stress whereas ${\rm Pe}<0$ demonstrates
 extensile stress.
 The splay and bend instability is shown by blue and red lines,
 respectively.
 The dashed lines show stability of the bulk system with $L=10$ in
 (\ref{bulk.stability}).
 The black lines indicate $\xi=\pm1$ and ${\rm Pe}=0$.
\label{fig.stability}
}
\end{center}
\end{figure}

Owing to the effect of rotational flow and shear alignment, the determinant of the linearized
matrix (\ref{linearised.eq.matrix}) becomes zero when the P\'{e}clet number becomes the critical P\'{e}clet
number, ${\rm Pe}_c$.
 This instability occurs both for $a_n^1$ and $b_n^1$, suggesting that
 both splay and bend instability is induced by this effect.
The physical meaning of the effect of the dipolar flow induced by the
stationary polar field  is sketched in Fig.~\ref{fig.isolated.shear.alignment}.
When the active stress is contractile, the orientation along the
$x$-axis is unstable, and a small perturbation of the first mode $m=1$
in Fig.~\ref{fig.psi.expansion} will grow.
The extensile active stress exhibits an opposite flow, and stabilizes
orientation along the $x$-axis. 
The flow generated by the perturbed polarity field also gives rise to
perturbation of the first mode as (\ref{kappa.m1}) and (\ref{omega.m1}).
The stability of the uniform polarity field is shown in
Fig.~\ref{fig.stability}.
This is obtained by zeroth of the determinant of
(\ref{linearised.eq.matrix}) with (\ref{stationary.kappa}),
(\ref{kappa.m1}), and (\ref{omega.m1}).
We truncate the Zernike expansion of $r$ in (\ref{psi.Zernike}) at the
lowest order necessary to find the transition.
The advection term is found to be small and is neglected. 
As activity increases, instability occurs both for contractile and
extensile stress in the rod-like flow-alignment regime ($\xi>1$).

We may also consider the system without boundary.
In this case,
(\ref{active.polar2.drop.model.normalise})-(\ref{active.polar2.drop.model.normalise.divv})
are linearized around the stationary state ${\bf P}^*= \hat{\bf x}$, that
is, $\psi^*=0$.
Note that in contrast to the active drop, there is no flow ${\bf v}^*=0$
at the stationary state of the bulk system.
In the Fourier space, the system becomes unstable when
\begin{align}
 - \frac{\zeta L^2}{2 \eta K}
 \cos 2\theta_k \left(
1 + \xi \cos 2\theta_k
 \right)
 > 1
 \label{bulk.stability}
\end{align}
where $\theta_k$ is the angle of the wave vector ${\bf k}$, and $L$ is
the system size.
The bend instability corresponds to $\theta_k=0$ whereas the splay
instability corresponds to $\theta_k=\pi/2$.
The result is shown in Fig.~\ref{fig.stability}.
Similar to the nematic case\cite{Edwards:2009}, the splay instability occurs when
$\xi <1$ for the contractile stress whereas the bend instability occurs
when $\xi>-1$.
The mechanism of this asymmetry between contractile and extensile stress
is that the rotational flow destabilizes the uniform orientation both by
contractile (splay instability) and extensile (bend instability) stress, whereas the shear flow
in the rod-like flow-alignment regime suppresses the instability only for the
contractile stress.
Our result suggests that the effect of the boundary enhances the
instability for the contractile stress for $\xi>1$.

\section{Chaotic Motion and Topological Defects}

\begin{figure}[h]
\begin{center}
\includegraphics[width=0.50\textwidth]{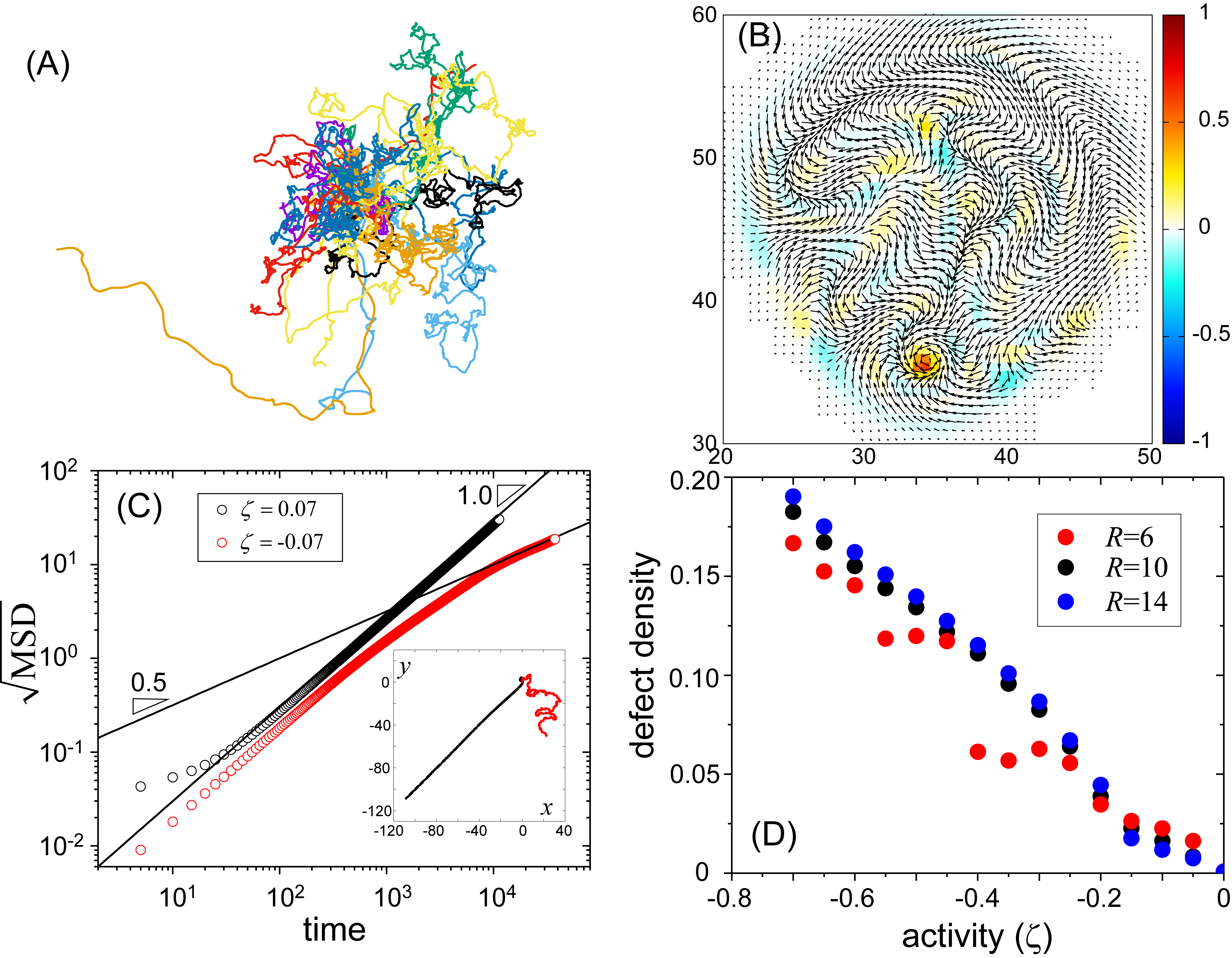}
\caption{ (Color Online) (A) Trajectories of chaotic motion. (B) The
 polarity field, ${\bf P}$, and defect density, $\rho$, of the active
 drop with $R=14$ and $\zeta=-0.2$ when it moves diffusively. The color
 of the defect density indicates topological charge.
When $\rho \simeq 0$, there is no defect.
 (C) Log-log plot of root mean-square displacement (MSD) of the active drop with $R=14$ for
 contractile ($\zeta =0.07$) and extensile ($\zeta=-0.07$) stress.
 The two lines indicate slope 0.5 and 1.0.
 Trajectories are shown in the inset.
 (D) Density of defects as a function of activity.
\label{fig.chaotic}
}
\end{center}
\end{figure}

When the activity is high, the self-propulsive motion is no longer periodic.
Its long-time behavior is diffusive whereas the short-time behavior is
ballistic, as shown in root mean-square
displacement $\sqrt{\rm MSD} = \sqrt{\langle |{\bf x}(t) - {\bf x}(0)|^2
\rangle}$ in
Fig.~\ref{fig.chaotic}(C).
During the diffusive motion, distortion of the polar field is accumulated
at lines, which are dynamically deforming (Fig.~\ref{fig.chaotic}(B)).
The shape of the drop is also fluctuating correlated with the motion of the line defects.
As activity is increased, the density of the line defects increases, and
at high extensile activity, disclinations appear.

In the case of an active {\it nematic} drop, the polarity field around a topological defect is expressed by
\begin{align}
 {\bf P}
 &=
 \left(
\cos \pm \frac{\theta}{2}, \sin \pm \frac{\theta}{2} 
 \right)
 ,
\end{align}
where $ \theta=\tan^{-1} y/x$.
The plus and minus signs denote $+1/2$ and $-1/2$ disclinations,
respectively.
The active stress is then given by
\begin{align}
 \sigma^{(a)}
 &=
 \frac{1}{2}
 \begin{pmatrix}
  \cos \theta & \pm \sin \theta \\
  \pm \sin \theta & - \cos \theta
 \end{pmatrix}
 ,
 \label{defect.nematic}
\end{align}
where we subtract the trace of the active stress because it merely
modifies pressure in incompressible systems.

The force dipole in (\ref{force.mulitpole}) generated by the active
stress can be obtained by integrating (\ref{defect.nematic}) inside the
drop.
 This clearly vanishes as $F_{ij}=0$.
The source dipole of $+1/2$ disclination is from (\ref{source.dipole}) 
\begin{align}
 q_i
 &=
 \zeta \pi
 \begin{pmatrix}
  1 \\ 0
 \end{pmatrix}
\end{align}
and for the $-1/2$ defect the source dipole vanishes ${\bf q}=0$.
As the source dipole is associated with the self-propulsive velocity,
the $+1/2$ disclinations move whereas $-1/2$ disclinations do not.
The same conclusion was reached by directly solving the Stokes equation in
\cite{Pismen:2013,Giomi:2014}.
Our analysis decomposes the self-propulsion into multipoles of forces and
associated flow.

In the case of defects of the polar field, the field is expressed as
\begin{align}
 {\bf P}
 &=
 \left(
\cos \theta, \pm \sin \theta
 \right).
\end{align}
After a calculation similar to that mentioned previously, we find there is neither
force dipole nor source dipole for the defects.

The position of the topological defect is extracted by the method in \cite{Mazenko:1997}.
The signed defect density is obtained as
\begin{align}
 \rho
 &=
 \frac{1}{2}
 \left(
 (\partial_x p_x) (\partial_y p_y)
 -  (\partial_x p_y) (\partial_y p_x)
 \right).
\end{align}
At the position of defects with positive topological charge $\rho \gg 0$, whereas for
negative topological charge $\rho \ll 0$.
The number density of the defects with positive and negative charges is
statistically the same.  
In Fig.~\ref{fig.chaotic}(D), the density of the positive defects is shown.
 As activity increases, the density linearly increases.

\section{Discussions and Summary}

In this work, we have analyzed the active polar drop, and showed that it
exhibits different type of self-propulsion depending on whether stress is contractile or
extensile.
The contractile stress results in translational motion, and, at higher
activity, rotational and zigzag motion.
In addition to these motions, chaotic diffusive motion occurs only with
extensile active stress.
The origin of the complex motion seems to be turbulence-like
behaviors in bulk owing to the active stress, which has been discussed in
the system without boundaries or with solid boundaries.
On the other hand, the self-propulsion under contractile stress has
a different origin: the flow generated by a surface force similar
to inhomogeneous surface tension in the Marangoni effect.
This extra effect, only existing in the confined drop, gives rise to
instability of the uniform polarity field.
This is in contrast to bulk systems where the contractile active
stress of rod-like molecules does not linearly destabilize the uniform
polar field. 

Our result of asymmetry between the contractile and extensile stresses
has some similarities to the results in \cite{Ramaswamy:2016}, in
which a contractile active polar fluid with frictional boundaries exhibits spontaneous flow and
oscillatory dynamics in the flow-alignment regime, whereas spontaneous flow,
oscillatory dynamics, and chaotic flow in the flow-tumbling regime.
Although there is no interfacial effect in that work and they studied
two different systems with different flow-alignment parameters, not
activity, they observed chaotic flow only in one regime.
It would be interesting as a future work to study nonlinear effects and
the mechanism of chaotic
flow.

Oscillatory dynamics such as traveling waves do appear as a
secondary bifurcation, not as a linear instability through Hopf
bifurcation.
This is in contrast to the observations of an active nematic
fluid\cite{Giomi:2011,Giomi:2012a} and active polar fluid with additional active stress in the
form of $\sigma^{(a)}_{ij} \sim \zeta_2 (\partial_i p_j + \partial_j
p_i)$ (see \cite{giomi:2008}).
In these cases, the systems have additional time scales associated with
the dynamics of the concentration field and/or velocity field (inertia
term).
Our system is in the low-Reynolds-number regime, and in the uniform
concentration and thus lacks those time scales.
Nevertheless, our system shows oscillatory dynamics such as zigzag
motion after several transitions from the stationary state.

 Recently, several experimental systems have been
 proposed to study self-propulsion of a drop containing liquid crystals
 with certain activity\cite{thakur:2006,Krueger:2016,Yamamoto:2017}.
Our model is probably too crude to explain the motion in these works,
but we hope to convey a basic understanding of these phenomena.

\begin{acknowledgments}
The authors are grateful to Rhoda Hawkins and Igor Aranson for
 helpful discussions.
 The authors acknowledge the support by JSPS KAKENHI Grant Nos. JP16H00793
 and 17K05605.
\end{acknowledgments}



\appendix

\section{Zernike polynomials}

Several lowest-order terms in Zernike polynomials are given by
\begin{align}
 R_1^1
 &=r
 \\
 R_3^1
 &=
 -2 r+3 r^3
 \\
 R_5^1
 &=
 3 r - 12 r^3 + 10 r^5
 \\
 R_7^1
 &=
 -4r + 30 r^3 -60 r^5 + 35 r^7
\end{align}
Orthogonality is expressed by
\begin{align}
 \int_0^1
R^m_n(r) R^{m}_{n'}(r)
 r dr
 &=
\frac{1}{2n+2}
 \delta_{n n'}
 \label{Zernike.orthogonality}
 .
\end{align}
The Laplacian acting on the Zernike polynomials results in\cite{Janssen:2014}
\begin{align}
&
 \Delta \left(
R_n^m \cos m\theta
 \right)
\nonumber \\
 =&
 \sum_{k=m,n-2}
(k+1)(n+k+2)(n-k)
 \left(
R_k^m \cos m\theta
 \right)
 .
 \label{Zernike.Laplace}
\end{align}
This operator makes sense when the mapping is $N+2$ dimensions onto
$N$ dimensions.
Note that $R_n^m=0$ for odd $n-m$ and therefore the Laplacian operator
induces the mapping from $[(N+2)/2]$ to $[N/2]$.

\section{Stability in Dynamics}
\label{sec.eigenvalues}

The analysis in the previous section is associated with the stability of
the steady state.
To see this, the time evolution of the linearized equation is expressed,
similar to (\ref{linearise.eq}), as
\begin{align}
 {\rm Pe} \partial_t \psi_1^{(1)}
 &=
 \Delta \psi_1^{(1)}
 - {\rm Pe} \mathcal{N} \psi_1^{(1)}
 .
 \label{linearise.eq.time}
\end{align}
Together with the boundary condition, we use the ansatz of $a_n^1(t)=
a_n^1 e^{\sigma t}$ and $b_n^1(t)= b_n^1 e^{\sigma t}$ as 
\begin{align}
 \Lambda
  \begin{pmatrix}
  a_n^1 \\
  b_n^1
 \end{pmatrix}
 &=
   \begin{pmatrix}
  L^{(aa)} & L^{(ab)} \\
    L^{(ba)} & L^{(bb)} 
 \end{pmatrix}
 \begin{pmatrix}
  a_n^1 \\
  b_n^1
 \end{pmatrix}
 ,
\end{align}
where
\begin{align}
 \Lambda
 &=
    \begin{pmatrix}
  \Lambda^{(aa)} & 0 \\
    0 & \Lambda^{(bb)} 
    \end{pmatrix}
 \\
 \Lambda^{(aa)}
 = \Lambda^{(bb)}
 &=
   \begin{pmatrix}
    0 & 0 & 0 & 0 & \cdots \\
    \sigma & 0 & 0 & 0 & \cdots \\
    0 & \sigma & 0 & 0 & \cdots \\
    0 & 0 & \sigma & 0 & \cdots \\
    \vdots & \vdots & \vdots & \vdots & \ddots
   \end{pmatrix}
 .
\end{align}
From $\det (L -\Lambda)=0$, the stability of the linearized equation is
obtained from $\sigma$.
When we neglect the second term on the right-hand side of (\ref{linearise.eq.time}),
all the $\sigma$ are negative, 
 and this suggests that the system relaxes to uniform orientation
without the coupling to the shear and rotational flow.

\section{Velocity Field under Force}
\label{sec.vel}

We solve the Stokes equation in two dimensions in the form of 
\begin{align}
 \Delta {\bf v} - \nabla p
 + {\bf f}
 &= 0
 \\
 \nabla \cdot {\bf v}
 &= 0
 .
\end{align}
We will consider only the first mode, which is expressed by $m=1$ in (\ref{vel.expansion}).
The pressure is given by
\begin{align}
 p
 &=
 {\bf p}_r \cdot {\bf n}
 .
\end{align}
With this expansion, incompressibility implies
\begin{align}
 {\bf v}_r' + \frac{{\bf v}_r}{r} - \frac{{\bf v}_t}{r}
 &= 0
 .
\end{align}
The Stokes equation is then rewritten as
\begin{align}
 {\bf v}_r'' + \frac{{\bf v}_r'}{r}
 - \frac{2 {\bf v}_r}{r^2}
 + \frac{2 {\bf v}_t}{r^2}
 - {\bf p}_r'
 + {\bf f}_r
 &=0
 \\
  {\bf v}_t'' + \frac{{\bf v}_t'}{r}
 - \frac{2 {\bf v}_t}{r^2}
 + \frac{2 {\bf v}_r}{r^2}
 - \frac{{\bf p}_r}{r}
 + {\bf f}_t
 &=0
 \\
 {\bf p}_r''
 + \frac{{\bf p}_r'}{r}
 - \frac{{\bf p}_r}{r^2}
 - {\bf g}(r)
 &= 0
\end{align}
where
\begin{align}
 {\bf g}(r)
 &=
 {\bf f}_r'
 + \frac{{\bf f}_r}{r}
 - \frac{{\bf f}_t}{r}
 .
\end{align}
The solution of the equation of the pressure is
 \begin{align}
 {\bf p}_r
 &=
 {\bf A} r
 \end{align}
 with the integral constant ${\bf A}$ to be determined by the boundary conditions.
The solution of the velocity field is obtained as
\begin{align}
 {\bf v}_r
 &=
 {\bf B} + \frac{{\bf A}}{8} r^2
 + {\bf h}_n (r)
 \\
 {\bf v}_t
 &=
 {\bf B}
 + \frac{3}{8} {\bf A} r^2
 + {\bf h}_t (r)
 \\
 {\bf h}_r
 &=
 \mp
  \int_0^r dr_1 \frac{1}{r_1^3}
 \int_0^{r_1} d r_2 \frac{r_2^2}{2}
 \odiff{}{r_2}(r_2 \bm{\psi}_1^{(1)})
 \\
 {\bf h}_t
 &=
 {\bf h}_r
 + r  {\bf h}'_r
\end{align}
 where ${\bf B}$ is another integral constant to be determined from the
 boundary condition, and  $\bm{\psi}_1^{(1)}=(\tilde{\psi}_1^{(1)}, \psi_1^{(1)} )$.
From the boundary condition, $( {\bf v}_r(R) - {\bf u}) \cdot {\bf
 n}=0$.
Using the self-propulsive velocity ${\bf u}$, the tangential velocity at
 the surface is $({\bf v}_t - {\bf u}) \cdot {\bf t} = v_s$, where
 the slip velocity, $v_s$ is associated with self-propulsive velocity
 ${\bf u} = -(1/2\pi) \int v_s {\bf t} d \theta$\cite{Yoshinaga:2018c}.
 The velocity field is then expressed as
 \begin{align}
  {\bf v}_r
  &=
  {\bf u}
  + \frac{\bf A}{8} (r^2 -R^2)
  + {\bf h}_r (r) - {\bf h}_r(R)
  \\
  {\bf v}_t
  &=
  {\bf u}
  +
  \frac{\bf A}{8} (3r^2 -R^2)
  + {\bf h}_t(r) - {\bf h}_r(R)
  \\
  {\bf u}
  &=
  -\frac{1}{8}
  \left(
  {\bf A} R^2
  + 4 \left(
  {\bf h}_t(R) - {\bf h}_n(R)
  \right)
  \right)
  .
 \end{align}

 Once the tangential slip velocity is known, we may calculate the
 velocity field outside the drop, and, assuming same viscosity outside
 the drop,  
the boundary condition (\ref{BC.stress}) results in
 \begin{align}
  {\bf A}
  &=
  \frac{2}{R^2}
  \left(
  {\bf h}_r (R) + R {\bf h}'_t(R)
  - {\bf h}_t(R)
  \right)
  \nonumber \\
  &=
  \frac{2}{R}
  \left(
{\bf h}'_r (R) - {\bf h}'_t(R)
  \right)
  .
 \end{align}
 
 In order to calculate (\ref{kappa.m1})and (\ref{omega.m1}), it is
suffice to compute ${\bf v}'_n + {\bf v}'_t$.
Using incompressibility and the result of the general solution, we obtain
\begin{align}
 {\bf v}'_r + {\bf v}'_t
 &=
 {\bf A} r
 \mp \frac{1}{2}
 \odiff{}{r}
\left(
 r \bm{\psi}_1^{(1)}
 \right)
 .
\end{align}
Using the recurrence relation of the Zernike polynomials
 \begin{align}
\odiff{}{r}
  \left(
r R_n^1(r)
  \right)
  =&
  2 R_n^1(r)
  + \frac{(n-1)(1+n(2 r^2-1))}{2n(r^2-1)} R_n^1(r)
  \nonumber \\
  &
  - \frac{(n+1)(n-1)}{2 n (r^2-1)} R_{n-2}^1(r)
  .
 \end{align}

\end{document}